\documentclass[twocolumn,showpacs,preprintnumbers,amsmath,amssymb]{revtex4-1}

\usepackage{graphicx}
\usepackage{dcolumn}
\usepackage{bm}
\usepackage{color}
\begin{document}

\title{Effect of Image Potential on Spin Polarized Transport through Magnetic Tunnel Junctions}

\author{Tehseen Z. Raza}
\affiliation{National University of Computer and Emerging Sciences, Islamabad, Pakistan}%

\author{Hassan Raza}
\affiliation{Center for Fundamental Research, Islamabad, Pakistan}%

\begin{abstract}
{We study the effect of image potential on spin polarized transport through Fe/MgO/Fe magnetic tunnel junctions in the presence of symmetry filtering. The image potential is included within the Simmon's model coupled with the non-equilibrium Green's function formalism to calculate the quantum transport. The increase in the current densities for the $\Delta_1$ symmetry and the $\Delta_5$ symmetry bands due to the image potential is more pronounced at higher bias, whereas, the increase in the magnitude of the tunnel magnetoresistance ratio is more prominent at lower bias for various barrier thicknesses.}
\end{abstract}


\maketitle

\section{INTRODUCTION}

Quantum mechanical tunneling is an important phenomenon not only for fundamental research, but also for various applications. Recently, the unique symmetry filtering property of MgO, where it preferentially tunnels the $\Delta_1$-like orbital symmetry states, has enabled various applications in memory and sensors. The use of MgO in Fe/MgO/Fe magnetic tunnel junction (MTJ) heterostructures has led to very high tunnel magnetoresistance (TMR) ratios due to the half-metallic $\Delta_1$ band in [100] direction for Fe. \%TMR ratios for Fe/MgO/Fe magnetic heterostructures have been theoretically predicted to be in excess of thousands \cite{Mathon01,Butler2001}, which were followed by experimental observations of about $200\%$ in CoFe/MgO/CoFe and fully epitaxial Fe/MgO/Fe MTJ devices at room temperature \cite{Yuasa04,Parkin04}.  
 
\begin{figure}
\vspace{2.8in}
\hskip -3.5in\includegraphics{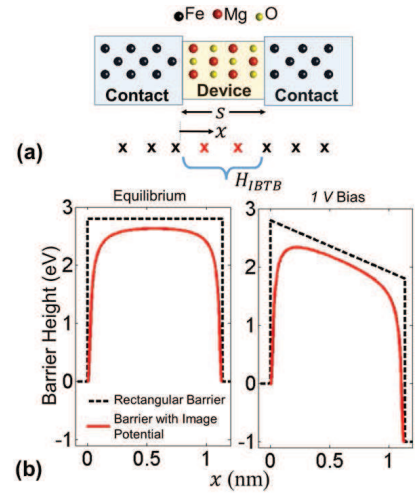}
\caption{(color online) Schematic device structure and image potential profile for a 4-layer device. (a) Ball and stick model for an Fe/MgO/Fe MTJ device and the corresponding lattice. The atomic visualization is done by using H\"uckel-NV \cite{HuckelNV}. (b) At equilibrium, the image potential reduces the barrier height and thickness for an otherwise rectangular barrier for a 4-layer device. With an applied bias, the barrier is further reduced due to the linear screening potential across the insulating barrier.}
\end{figure}

Image potential plays an important role in metal based tunneling geometries. In this context, Simmon's model  \cite{SimmonsSim63, SimmonsDissim63} has been widely used in tunneling calculations. Within this model, the image potential not only lowers the tunneling barrier but also reduces the effective thickness by modulating the barrier shape, thereby affecting the transport. In MgO based systems, where the symmetry filtering governs the transport properties, the effect of image potential on the non-equilibrium transport may not be established a priori. In this paper, our motivation is to study how the spin polarized quantum mechanical tunneling is affected by the image potential through Fe/MgO/Fe MTJs in the presence of symmetry filtering. 

This paper has been organized into four sections. We discuss the theoretical model in Sec. II, followed by the results in Sec. III. Finally, we provide the conclusions in Sec. IV. 

\section{THEORETICAL MODEL}

\begin{table}[!t]
\renewcommand{\arraystretch}{1.6}
\caption{\label{tab:Table1} Spin-polarized IBTB parameters (in eV). Band offset ($E_{bo}$) and hopping parameter ($t_{o}$) for the majority($\uparrow$) spin band and the minority($\downarrow$) spin band of bcc-Fe(100) are reported \cite{RazaTNANO}.}
\begin{tabular}{llcccc}
\hline\hline
Band           &Symmetry                                &$t_{o}$&                   & $E_{bo}$\\ \hline
               &                                     &$\uparrow$  &  $\downarrow$       &  $\uparrow$  &  $\downarrow$ \\
\hline\hline
$\Delta_{1}$   & $4s, 4p_{z}, 3d_{z^{2}}$            & 2.5         &  2.5                & -1             & 1\\
$\Delta_{5}$   & $4p_{x}, 4p_{y}, 3d_{xz}, 3d_{yz}$  & 1           &  1                  & -3.5           & -2.0\\
$\Delta_{2}$   & $3d_{x^{2}-y^{2}}$                  & -0.2        &-0.35                & -2.1           & -0.8\\
$\Delta_{2'}$  & $3d_{xy}$                           & 0.2         &  0.2                & -1.5           & 0.4\\
\hline
\end{tabular}
\end{table} 

\begin{figure}
\vspace{2.25in}
\hskip -3.8in\includegraphics{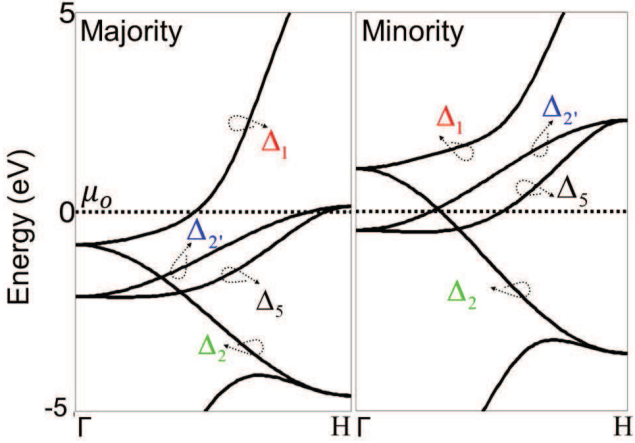}
\caption{(color online) Symmetry bands for bcc-Fe(100). $\Delta_{1}$, $\Delta_{5}$, $\Delta_{2}$ and $\Delta_{2'}$ bands for the majority spin and the minority spin by using extended H\"uckel theory \cite{Cerda2000}.}
\end{figure}

The device structure is shown in Fig. 1(a) with the corresponding lattice. We use independent-band tight-binding (IBTB) model coupled with the Non-Equilibrium Green's Function (NEGF) formalism to calculate the spin-polarized quantum transport \cite{RazaTNANO} through these magnetic heterostructures. In IBTB method, each lattice point corresponds to a unit cell. Thus, for a 4-layer device shown in Fig. 1(a) there are two lattice points corresponding to the two MgO unit cells. The energy band diagram for a heterostructure with insulating region is shown by a rectangular barrier in Fig. 1 (b). Under equilibrium, the image potential modulates the barrier shape, effectively lowering the barrier height and reducing the barrier thickness. The effect of image potential ($U_I$) is included within the Simmon's model as follows \cite{SimmonsSim63}:

\begin{eqnarray} U_I = -1.15 \lambda \frac{s^2}{x(s-x)} \end{eqnarray}

where \begin{eqnarray} \lambda = e^2 \frac{ln2}{8\pi\epsilon s} \end{eqnarray}

In the above equation, $s$ is the barrier thickness, $x$ is the distance from the left electrode as shown in Fig. 1(a), $\epsilon$ = 11.9$\epsilon_o$ is the bulk permittivity of MgO and $e$ is the electronic charge. Typical experimentally investigated thicknesses are between 6 and 15 monolayers of MgO \cite{Yuasa04}. According to the atomic structure for crystalline Fe/MgO/Fe \cite{Heiliger05}, after including the atomic relaxation at the interface, the barrier thicknesses for the 4-layer, 8-layer and 12-layer Fe/MgO/Fe devices are about \textit{1.13\ nm}, \textit{1.99\ nm} and \textit{2.85\ nm}, respectively. 
 
In order to calculate the transmission for homogeneous materials by using the IBTB method, we start with the following Hamiltonian for each symmetry band \cite{Kittel_Book},
\begin{eqnarray}H_{IBTB}=\begin{cases}E_{bo}+2t_o\ \ \ for\ i=j\cr -t_o\ \ \ \ \ \ \ \ \ \ for\ |i-j|=1 \end{cases}\end{eqnarray}
where $E_{bo}$ is the band offset and $t_o$ is the hopping parameter. This results in a cosine dispersion as follows:
$\epsilon(k)=E_{bo}+2t_{o}[1-\cos(ka_{l})]$, where  $k$ is the lattice wave vector and $a_{l}$ is the lattice spacing in the transport direction. The IBTB hopping parameters and the band offsets for various bcc-Fe(100) symmetry bands are reported in Table I \cite{RazaTNANO}. The Fe symmetry bands for the majority spin and the minority spin are also shown in Fig. 2 which are calculated by using extended H\"uckel theory \cite{Cerda2000}. 

The IBTB method has been benchmarked against \textit{ab initio} studies. Moreover, the IBTB parameters are transferable for various barrier thicknesses where a quantitative agreement in current densities and TMR ratios has been obtained. The model adequately captures the high-bias trends as well. 

For a heterostructure like Fe/MgO/Fe, the Hamiltonian in Eq. 1 is modified depending on the device material. \textit{e.g.}, in our system the device is made up of MgO tunnel barrier. We use $U_b$ and $t_o$ to specify the MgO barrier. These IBTB parameters for MgO are reported in Table II \cite{RazaTNANO}. At the heterostructure interface, the off-diagonal elements are chosen such that the resulting Hamiltonian is Hermitian. 

\begin{table}[!t]
\renewcommand{\arraystretch}{1.3}
\caption{IBTB parameters (in eV) for MgO. MgO barrier height $U_{b}$ and hopping parameter $t_{o}$ for the $\Delta_1$ symmetry band and $\Delta_5$ symmetry band are reported \cite{RazaTNANO}.}
\label{tab:Table2}
\centering
\begin{tabular}{lcc}
\hline\hline
               & \ \ \ $t_{o}$                 & \ \ \  $U_{b}$         \\
\hline\hline
$\Delta_{1}$ band             & 0.64                    &  2.8             \\
$\Delta_{5}$ band             & 0.64                    &  4.5             \\
\hline
\end{tabular}
\end{table}

\begin{figure}
\vspace{2.6in}
\hskip -3.8in\includegraphics{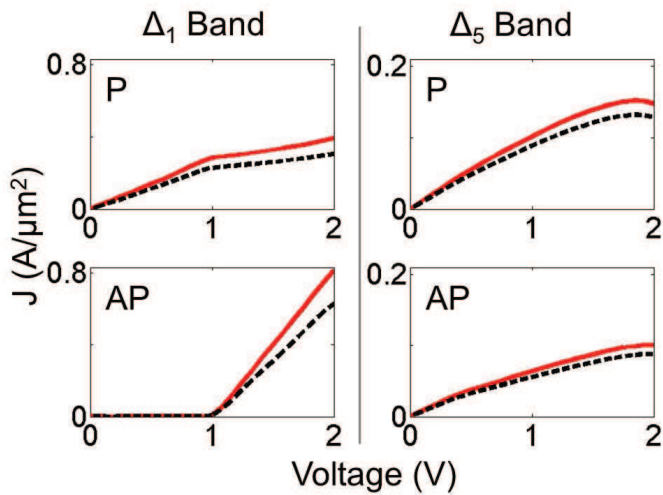}
\caption{(color online) Transport through $\Delta_1$ band and $\Delta_5$ band. The effect of image potential on the P and AP current densities for a 4-layer Fe/MgO/Fe MTJ device. The current densities with and without image potential are shown by solid (red) and dotted (black) lines, respectively.}
\end{figure}

\begin{figure*}
\vspace{2.6in}
\hskip -3.8in\includegraphics{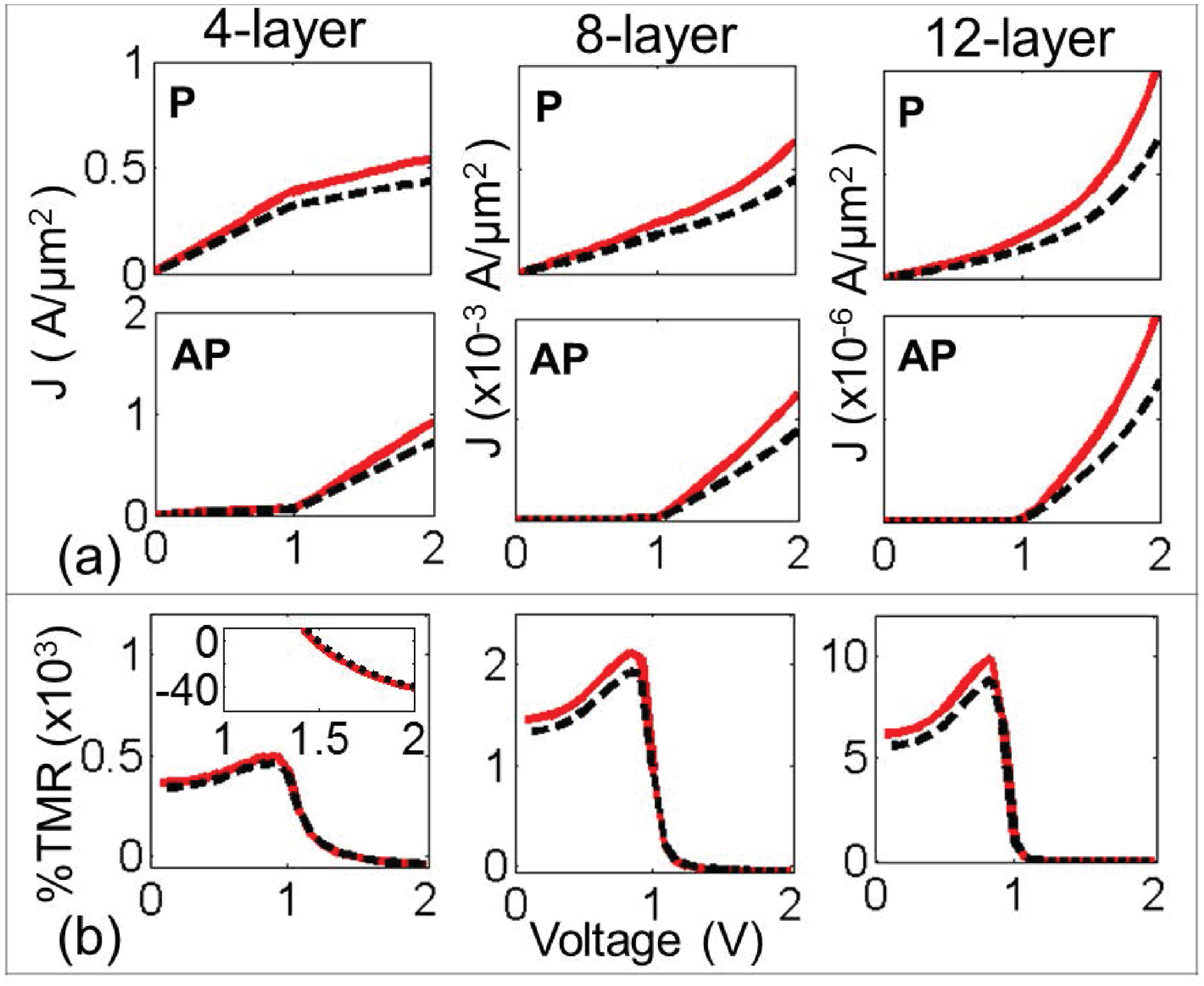}
\caption{(color online) Transport for various barrier thicknesses. (a) P and AP current densities for 4-layer, 8-layer and 12-layer devices with image potential (solid red line) and without image potential (dotted black line). The effect of image potential is more pronounced at higher bias. (b) Optimistic \%TMR. The effect of image potential is more pronounced at lower bias for 4-layer, 8-layer and 12-layer devices.}
\end{figure*}

We use linear screening for the Laplace potential in the tunnel barrier region. The image potential ($U_I$) and the Laplace potential ($U_L$) are included in the Hamiltonian matrix as follows: 

\begin{eqnarray} H_{IBTB}=\begin{cases}E_{bo}+2t_o+U_L+U_I\ \ \ for\ i=j\cr -t_o\ \ \ \ \ \ \ \ \ \ \ \ \ \ \ \ \ \ \ \ \ \ \ \ \ for\ |i-j|=1\cr 0\ \ \ \ \ \ \ \ \ \ \ \ \ \ \ \ \ \ \ \ \ \ \ \ \ \ \ \ otherwise \end{cases}\end{eqnarray}

The Green's function for the orthogonal IBTB is then calculated as:
\begin{eqnarray}\hat{G}=[(E_l+i0^+)I-H_{IBTB}-\Sigma_{1,2}]^{-1}\end{eqnarray}
where $\hat{\Sigma_{1,2}}=-t_oe^{ik_{1,2}a_{l}}$ are the self-energies for the left and right contact respectively, $E_l$ is the longitudinal energy along the device and $I$ is the identity matrix.   

We further use NEGF to calculate the transmission. Assuming an infinite device cross-section, for each $k_{||} = (k_x,k_y)$ in the transverse Brillouin zone (BZ), the transmission in the longitudinal direction is given as $\hat{T}(E_l)=tr(\hat{\Gamma_{1}}\hat{G}\hat{\Gamma_{2}}\hat{G^{\dagger}})$, where $\Gamma_{1,2}=i(\hat{\Sigma_{1,2}} - \hat{\Sigma_{1,2}}^{\dagger})$ are the contact broadening functions.   

Finally, the transmission per unit area is calculated analytically by summing over the 2D transverse BZ as follows: 
\begin{eqnarray}T_{IBTB}=\frac{1}{4\pi^2}\int_{(-\frac{\pi}{a},-\frac{\pi}{a})}^{(\frac{\pi}{a},\frac{\pi}{a})}dk_{y}dk_{z}\hat{T}(E_l)=\frac{1}{a^2}\hat{T}(E_l)\end{eqnarray} 
where $a=$ \textit{2.86 {\AA}} is the Fe cubic lattice constant.

The current densities in the parallel (P) and the anti-parallel (AP) configurations for the $\Delta_1$ band and the $\Delta_5$ band, per spin, are then calculated independently by using the Landauer’s formula as follows:
\begin{eqnarray}J=\frac{e}{h}\int{dE_{l} \ T_{IBTB} \ [f_1 - f_2] }\end{eqnarray} 
where $f_1$ and $f_2$ are the Fermi's functions for the left and right contacts respectively. For each band, the current in either the P or the AP configuration is made up of two components. In the absence of spin flip scattering, the current density in the P configuration, where both the Fe contacts have their magnetization aligned, is given by $J_P = J_{P\uparrow\uparrow} + J_{P\downarrow\downarrow}$. For the antiparallel configuration, the magnetizations of the two contacts are opposite, and hence $J_{AP} = J_{AP\uparrow\downarrow} + J_{AP\downarrow\uparrow}$. The total current density $J_{total}$ in P and AP configurations is then calculated by summing the current densities through the $\Delta_1$ and $\Delta_5$ bands. Only $\Delta_1$ and $\Delta_5$ bands are considered and $\Delta_2$ and $\Delta_{2'}$ bands are ignored due to their large decay rates \cite{Butler2001}. The optimistic tunnel magnetoresistance ratio (TMR) is finally calculated as:
\begin{eqnarray}TMR = \frac{J_{P\ \textit{total}}-J_{AP\ \textit{total}}}{J_{AP\ \textit{total}}}\end{eqnarray} 

\section{RESULTS AND DISCUSSION}

The effect of image potential on the transport through the individual bands is shown in Fig. 3. The image potential effectively lowers the barrier height and thickness thus increasing the current density through the device. For Fe/MgO/Fe device, as the tunneling probability of various symmetry bands is different due to the symmetry filtering property of MgO, the image potential lowers the effective barrier for various symmetry bands uniquely. Under the applied bias, the Laplace potential is assumed to drop linearly across the insulating barrier. Such linear screening is a reasonable assumption for insulating barriers. This however, leads to an additional barrier lowering and thinning, effectively changing a rectangular barrier into a triangular one. Thus, the current is incrementally higher with the applied bias. 

In P configuration for the $\Delta_1$ symmetry band, a threshold behavior is observed in the current density at about $1\ V$ with piecewise linear characteristics. Unlike the minority band edge for the $\Delta_1$ symmetry band at $1\ V$, the majority band edge at $-1\ eV$ leads to states around the chemical potential $\mu_o$. The total $J_P$ for this band is thus dominated by $J_{P\uparrow\uparrow}$ current density where both the contacts have majority spin configuration. At around $1\ V$ bias, the conducting states start to decrease, leading to a decreased slope in the P current density. On the other hand, the AP current density for the $\Delta_1$ symmetry band increases sharply after about $1\ V$. This is due to the half-metallic $\Delta_1$ band for bcc-Fe in [100] direction. With the increasing bias, the minority spin band is pulled down till it starts to contribute to states within the conduction window. 

Moreover, at high bias, the $\Delta_5$ band current density starts to decrease. This is expected, since the bandwidth of this band is $4\ eV$ with the minority spin band edge at about $2\ eV$. Therefore, at a bias of about $2\ V$ the available states at the Fermi energy start to decrease exhibiting a negative differential resistance (NDR) effect. One of the possible ways where such systems may exhibit NDR is if the insulating barrier is designed to predominantly facilitate the tunneling through the $\Delta_5$-like symmetry states. The analysis of such a system is left as a future direction. Another system through which similar NDR behavior has been reported is an electronic structure modulation transistor where graphene nanoribbon is used with a finite bandwidth mid-gap state \cite{HassanEMT}.

Apart from this, the total P current density is dominated by the $\Delta_1$ band at low bias due to a lower potential barrier for these symmetry states. On the other hand, the total AP current density is dominated by the $\Delta_5$ symmetry band for low bias. As mentioned earlier, after a critical voltage there is a sharp increase in the AP current of the $\Delta_1$ symmetry band. This increase leads to a sharp roll-off in TMR, which ultimately becomes negative when the AP current density overcomes the P current density.  

Fig. 4 shows the total P and AP current densities for 4-layer, 8-layer and 12-layer devices with the corresponding optimistic \%TMR. The current densities are clearly higher with the inclusion of image potential. Furthermore, the magnitude of \%TMR with the image potential is always greater than the magnitude of \%TMR without the image potential under all bias conditions. However, this effect is more pronounced at lower bias for various barrier thicknesses. The inset shows the magnified view graph for the bias where the \%TMR becomes negative for a 4-layer device. This bias threshold of \%TMR becoming negative is approximately $1.5 V$, $1.3 V$ and $1.2 V$ for 4-layer, 8-layer and 12-layer devices, respectively. 

Furthermore, the image potential effects are also strongly dependent on the dielectric constant of the insulator, since the potential varies hyperbolically with the dielectric constant. Although, we show that the image potential plays an important role in MgO based MTJ devices, these effects will become more pronounced for a device, where the barrier material has a lower dielectric constant. Yet another example, where the image potential effects are reported to be important due to lower dielectric constant, is the conduction through molecular junctions \cite{Raza08}. 

\section{CONCLUSIONS}

We have reported that incorporating image potential is vital in understanding the transport through Fe/MgO/Fe based tunneling heterostructures. Image potential modulates the barrier height and the barrier thickness in an intricate way. Combined with the symmetry filtering property of Fe/MgO/Fe based heterostructures, the effect becomes sensitive to the barrier details for each symmetry band. The current densities show a pronounced increase at high bias, whereas the magnitude of \%TMR shows a prominent increase at lower bias. 

\section*{ACKNOWLEDGEMENTS}

The authors would like to acknowledge Dr. Jorge Iribas Cerda for useful discussions.

\end{document}